# Meta-compilation of Baseline JIT Compilers with Druid

Nahuel Palumbo[a], Guillermo Polito[a], Stéphane Ducasse[a], and Pablo Tesone[a]

a Université de Lille, Inria, CNRS, Centrale Lille, UMR 9189 CRIStAL, Lille, France

**Abstract** Virtual Machines (VMs) combine interpreters and just-in-time (JIT) compiled code to achieve good performance. However, implementing different execution engines increases the cost of developing and maintaining such solutions. JIT compilers based on meta-compilation cope with these issues by automatically generating optimizing JIT compilers. This leaves open the question of how meta-compilation applies to baseline JIT compilers, which improve warmup times by trading off optimizations.

In this paper, we present Druid, an ahead-of-time automatic approach to generate baseline JIT compiler frontends from interpreters. Language developers guide meta-compilation by annotating interpreter code and using Druid's intrinsics. Druid targets the meta-compilation to an existing JIT compiler infrastructure to achieve good warm-up performance.

We applied Druid in the context of the Pharo programming language and evaluated it by comparing an autogenerated JIT compiler frontend against the one in production for more than 10 years. Our generated JIT compiler frontend is 2× faster on average than the interpreter and achieves on average 0.7× the performance of the handwritten JIT compiler. Our experiment only required changes in 60 call sites in the interpreter, showing that our solution makes language VMs **easier to maintain and evolve in the long run**.

## 1 Introduction

Language Virtual Machines (VMs) use multiple execution tiers to optimize programs at run time [4, 2]. They are often structured around an interpreter and one or more just-in-time (JIT) compilers to achieve a good balance between runtime, program analysis, and compilation time [17, 37, 12, 52].

In such architectures, programming language features must be implemented more than once, often in different ways, making these machines difficult to build, understand, and extend [21]. Reimplementing the same features at different levels of abstraction increases the risk of introducing bugs due to differences in the execution [42]. These facts increase the cost of language implementations and disrupt the tasks between language and VM developers (c.f., Section 2).

Meta-compilation approaches [52, 10, 53, 46] propose to reduce such complexity and risks by automatically generating compiled code from interpreters. These solutions are based on the first Futamura projection, applying partial evaluation to the interpreter for a given program to generate optimized machine code [23]. Language developers focus on developing a language interpreter, which is simpler to implement and debug than a compiler [21]. In these approaches, meta-compilation happens at run time, making them sensitive to higher compilation and warmup times.

In this article, we present Druid, an ahead-of-time meta-compiler that generates baseline JIT compiler code from interpreters, inspired by the second Futamura projection [23]. Our approach is not based on partial evaluation, but a hand-written meta-compiler (*Cogen*) [9] (c.f., Section 3). More concretely, we propose to target the code generation to an existing JIT compiler *backend* (c.f., Section 4), thus generating a compiler *frontend* [48]. Our solution introduces interpreter annotations and intrinsic functions language that developers use to guide meta-compilation. Moreover, Druid stages computations to be executed by the JIT compiler instead of run time.

This architecture allows the separation of concerns and maintenance of VM components. Language interpreters are defined by language implementors based on the interpreter framework while JIT compiler front ends are automatically generated ahead of time by the Druid meta-compiler.

In this paper, we study the following research questions:

**RQ1:** Does our solution generate JIT compilers that improve the performance of the interpreter?

**RQ2:** Is the JIT compiler generated by Druid competitive in performance against the handwritten version?

**RQ3:** Does our generated approach increase the overhead at compile time?

**RQ4:** How much development effort is affected by our automatic approach?

**Results** We evaluated our approach using the Pharo VM in production for over 20 years (c.f., Section 5). The Pharo VM is based on an interpreter and a JIT compiler handwritten by experts that we used as comparison baseline [37, 36].

Our results show that the JIT compiler generated by Druid is 2× faster than the interpreter on average, and 5× in our best scenario. Moreover, our prototype achieves up to 0.7× the performance of the handwritten JIT compiler front end engineered over 10 years. Finally, our autogenerated front end shows JIT compile times similar to the handwritten one (difference of 0.18 % on average).

**Contributions**

- A description of Druid, an ahead of time meta-compiler generating baseline JIT compiler frontends from interpreters.
- The changes introduced in the interpreter and JIT compiler architecture to implement our solution using the production Pharo VM.
- A comparison of runtime performance, development effort, and JIT compilation overhead between the autogenerated front end and the handwritten one.

## 2 Background: Translating Interpreters to Baseline JITs

This section introduces the basic concepts required throughout the paper.

### 2.1 Interpreters, Compilers and JIT Compilers

The goal of a compiler is to translate a program written in a source language to an equivalent program written in a target language. Usually, when comparing interpreted vs. compiled language implementations, such translation has the goal of reducing execution time, assuming that a program will be compiled once and executed many times. In other words, the goal is to satisfy the following equation.

$$target = compiler(source) \tag{1}$$

$$t_{interpreter}(source, input) > t_{target}(input) \tag{2}$$

Where $t_{interpreter}(source, input)$ represents the time interpreting *source* with input *input*, and $t_{target}(input)$ represents the time executing the compiled *source* in *target* language with input *input*.

JIT compilers take such constraints to the extreme because compilation happens at run time. The following equation introduces $t_{compiler}(source)$, which represents the time compiling *source* to a *target* language. When compilation time is considered at run time, the sum of the compilation time and the execution time should not be greater than the interpreter time for compilation to be worth.

$$t_{interpreter}(source, input) > t_{compiler}(source) + t_{target}(input) \tag{3}$$

JIT compilers are built around this compilation time vs. run time tradeoffs. Optimizing JIT compilers minimize $t_{target}$ at the cost of higher $t_{compiler}$ times. Baseline JIT compilers minimize $t_{compiler}$ at the cost of the optimizations applied. Most of the time, baseline JIT compilation removes interpretation overhead.

Optimizing and baseline JIT compilers are combined in practice to achieve the best of both worlds: lots of methods that are rarely executed are compiled with baseline JIT compilers, and few methods that are commonly executed are compiled with optimizing JIT compilers [27, 25].

### 2.2 VMs as Multi-staged Machines

A virtual machine can be seen as a multi-staged engine because it usually contains more than one execution engine with different properties [25, 27].

**Definition 1** (Interpreter). *An **interpreter** is a one-stage engine: it directly executes source code to get the result. It easily interacts with other runtime components, such as the memory manager and the garbage collector, because they are defined at the same level of abstraction.*

**Definition 2** (JIT compiler). *A **JIT compiler** is a two-stage engine: instead of executing source code, it generates residual (machine) code with the same behavior as the source code, that will later be executed at run time. Such translation requires, for example, understanding the transformation of code between two different representations/abstraction levels (*e.g.*, stack bytecode to register machine code), the difference between what can be executed at each stage (compile time vs. run time), and deciding the scope of the compilation* i.e., *when should the compiler stop and emit a runtime function call to* e.g., *run the GC.*

When in multi-staged machines, only one tier (typically the interpreter) must support the entire language implementation. Other tiers usually trade-off space by execution time, compiling only common paths and falling back to the interpreter when the rare paths are hit.

### 2.3 Meta-compilation Targetting Baseline JIT Compilers

In this paper, we study the meta-compilation of baseline JITs. As stated before, the goal of the baseline JIT compiler is to compile fast. To generate a baseline JIT compiler from an interpreter, meta-compilation can be engineered as a **one-stage to two-stage execution engine transformation**.

```
Interpreter >> bytecodeDup
  self push: self stackTop
```

```
JITFrontend >> gen_bytecodeDup
  self MoveMw: 0 r: SPReg R: TempReg.
  self PushR: TempReg
```

**Figure 1** Interpreter vs. Baseline JIT frontend for bytecode duplicate top.

To illustrate this transformation, Figure 1 shows an example of the bytecode that duplicates the top of the stack as defined in the Pharo virtual machine [29, 37]. The bytecode is defined both in the interpreter (left) and in the compiler frontend (right) using a dynamic machine code generator named Cogit [36]. Our solution, Druid, is a source-to-source meta-compiler for the Pharo VM that takes as input the Pharo interpreter (left of Figure 1) and outputs the domain-specific language used to implement the JIT compiler frontends for the Pharo VM (right of Figure 1).

At JIT compile time (*i.e.,* during program execution), if the bytecode *dup* is found in a stream of bytecodes, the Cogit JIT compiler executes the method `gen_bytecodeDup` to generate the intermediate representation of the method under compilation. Then, it generates machine code and executes it at run time to obtain a program's result.

The details of the solution are found in Section 3.

### 2.4 Challenges of Baseline JIT Generation from Interpreter Definition

Our meta-compilation approach translates interpreter code into JIT compiler code that implements the same semantics. However, JIT compilation can take advantage of the translation step to *weave* decisions into the generated code for optimization purposes. For example, not interpreter features are worth being JIT-compiled: compiling slow paths in the handler definition may occupy additional space and execute infrequently, thus the JIT compiler could emit a call to the interpreter instead. Also, the JIT compiler can take advantage of compilation invariants such as values that are JIT compile-time constants, and fold them.

The challenge is thus to embed such decisions into the generated code, without runtime information during meta-compilation. This presents the following challenges:

**Ahead-of-time compilation scoping.** *The baseline JIT meta-compiler needs to statically decide what language handler paths should be compiled.* A VM containing many execution engines must have at least one supporting the entire language specification. This is typically the role of an interpreter. On top of that, JIT compilers can decide to compile, for an instruction handler, a subset of all paths and fall back on the slower engines for other cases. This allows JIT compilers to focus on common or important language features. Although traditional optimizing JIT compilers drive those decisions using profiling information obtained at run time, an ahead-of-time meta-compiler should provide mechanisms to statically determine the slow paths inside the handler definition to avoid their (meta-)compilation.

**Switching between generated code and the runtime system.** *The baseline JIT meta-compiler needs to statically decide when to introduce runtime calls.* On the one hand, interpreters are typically implemented in the same programming language as the rest of the runtime (*e.g.,* the memory manager, the JIT compiler), and thus they have a seamless integration. On the other hand, JIT compilers need to explicitly manage the interactions between JIT-compiled code and the runtime environment *e.g.,* if JIT-compiled code invokes the GC, values stored in registers must be spilled in the stack if the GC traverses the stack to find roots.

**Different staging levels.** *The baseline JIT meta-compiler needs to statically decide whether operations should be executed ahead of time, JIT-compile time, or run time, producing a residual program that will execute code at different stages.* As said in Section 2.2, interpreters are one-stage machines that directly execute code. Compilers are two-stage machines where compilation decisions are taken statically and ahead of time. JIT compilers, instead, have runtime information *e.g.,* some properties and values are known constant at JIT compile time and can be used to drive optimizations.

### 2.5 Vocabulary: Bytecode and Primitive Methods

For clarity of the presentation, this subsection presents some vocabulary points. As a Smalltalk descendant, the Pharo programming language is implemented in terms of *bytecode* and *primitive* instructions [26].

**Bytecode instructions.** Bytecode instructions are used as a VM intermediate language. Pharo source code is compiled to a sequence of stack-based bytecode instructions *e.g.,* push instance variable, duplicate the top of the stack.

**Primitive Methods.** Primitive methods are normal Pharo methods that are annotated to call a primitive instruction (see below).

**Primitive instructions.** Primitive instructions are native operations exposed by the virtual machine, equivalent to native methods in other language implementations [2]. Primitive instructions implement basic functionality such as arithmetics and object allocation. By design, the primitive instructions can *fail* and activate a fall-back bytecode method. Otherwise, when a primitive instruction succeeds, the execution engine returns control to the caller with the result.

From the point of view of this paper, we will consider both bytecode and primitive instructions as VM *instructions*. Some functionality in our VM implementation is provided as both bytecode *and* primitive instructions, duplicated for performance reasons *e.g.,* integer addition. Moreover, primitive instructions tend to be larger and more complex than bytecode instructions [42].

**Instruction Handlers** In the PharoVM, both the interpreter and the JIT compiler front end are implemented using a table dispatch approach. Bytecode opcodes are mapped to functions called *handlers*. Handlers in the interpreter execute the instruction, while handlers in the JIT compiler frontend generate nodes in the JIT compiler intermediate representation (IR). In the rest of this paper, as a shortcut, we refer to the handlers in the JIT compiler frontend as *IR generators*. Figure 2 illustrates the handlers on the JIT compiler front end.

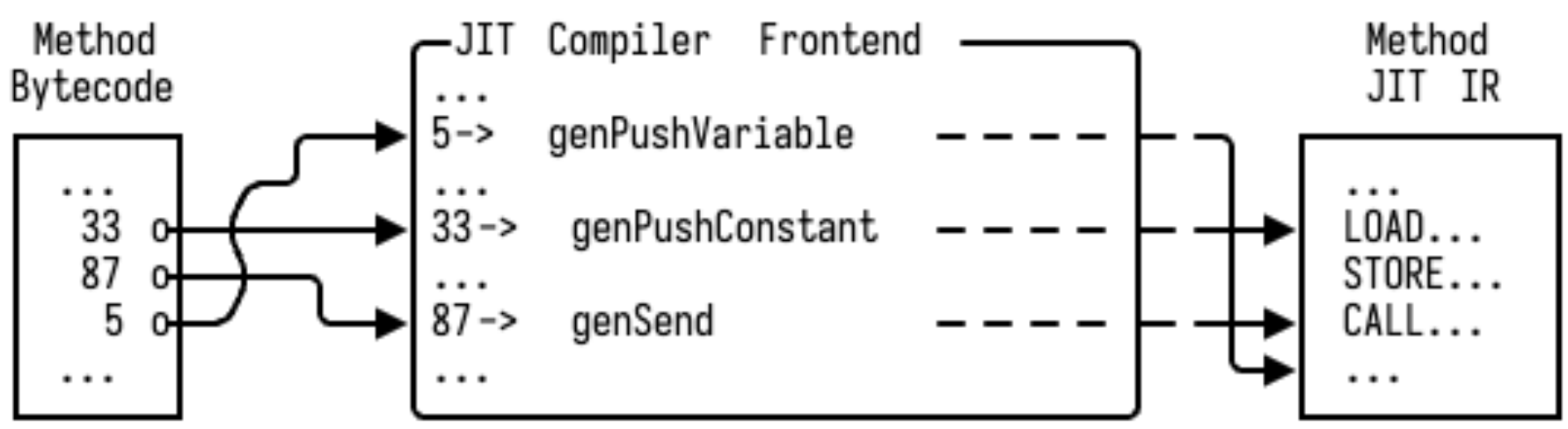

**Figure 2** The JIT compiler front end uses a table dispatch approach, mapping bytecode opcodes to handler functions generating intermediate representation nodes.

## 3 Druid: AoT Meta-compilation of a Baseline JIT Compiler

We built Druid,[1] an ahead-of-time *meta-compiler* that generates JIT compiler frontend code from a language interpreter. Conceptually, Druid works as a compiler generator program (Cogen) [9]. We opted for simplicity and control to implement our meta-compiler as a hand-written compiler instead of using a partial evaluator [23, 25]. To prototype these ideas, we extended the Pharo VM interpreter framework [29, 37] to solve the challenges expressed in Section 2.4:

**Intrinsics and annotations drive scoping and runtime interactions.** The interpreter code is annotated and extended with intrinsics to guide the JIT frontend generation.

**Staging separates JIT compile time vs run time.** A staging pass at the end of the compilation pipeline statically determines code that is executable at JIT-compile time.

### 3.1 Druid Meta-compilation Overview

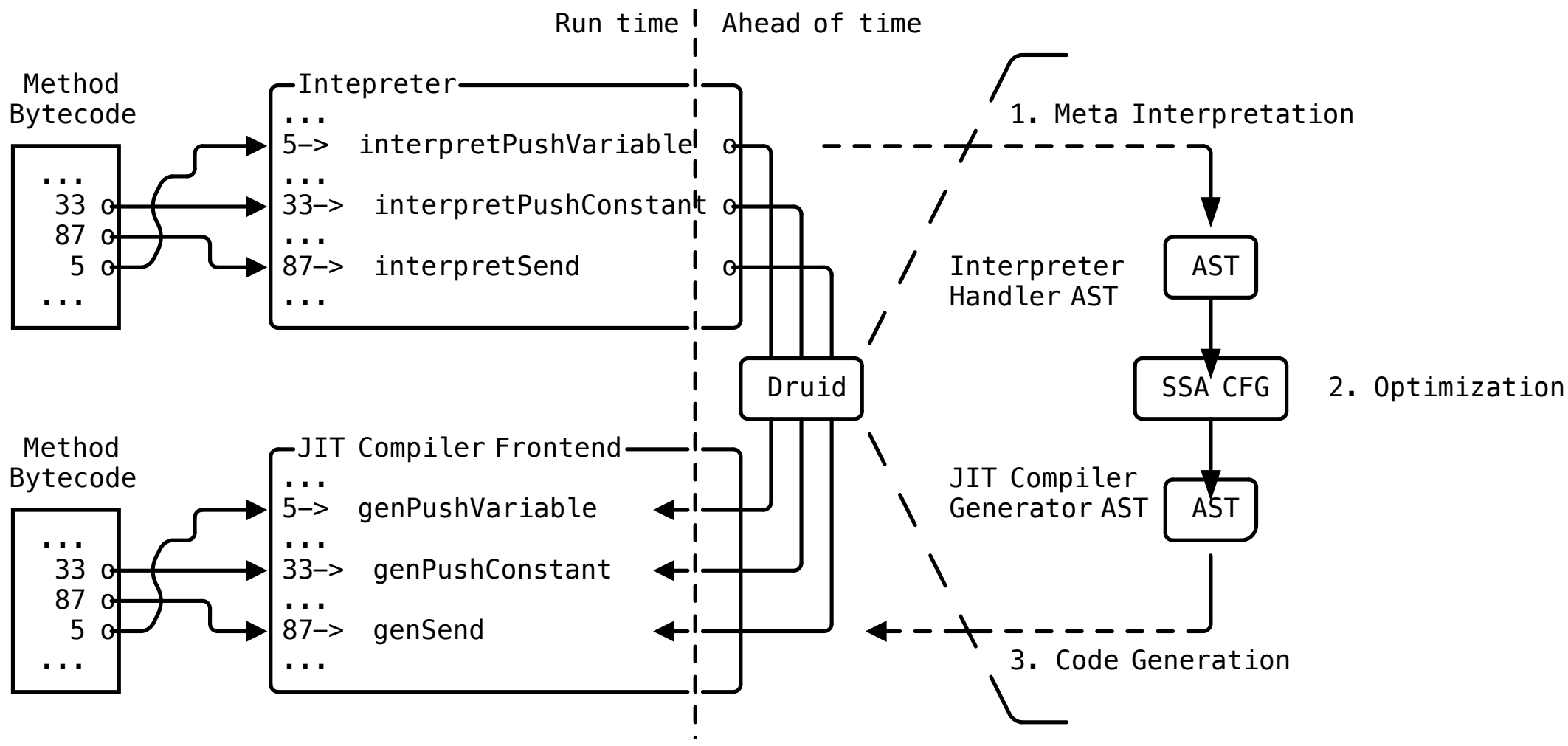

■ **Figure 3** Druid's meta-compiler architecture. Druid generates ahead of time a baseline JIT compiler frontend from an interpreter, translating interpreter bytecode handlers into IR generator handlers. Druid generates an SSA-form control flow graph IR and optimizes it using traditional algorithms. Finally, it translates this IR to a JIT compiler front end deployed along the rest of the VM source code.

As explained briefly in the previous section, Druid is a source-to-source ahead-of-time meta-compiler for the Pharo VM. We illustrate Druid in Figure 3. Druid's input is an interpreter and its output is the source code of its corresponding JIT compiler frontend. The Druid meta-compiler is engineered as a traditional multi-pass compiler with additional custom passes designed for meta-compilation.

[1] https://github.com/Alamvic/druid visited on 2024-02-04.

**Druid's front end** (Figure 3, step (1)) Druid's front end is a meta-interpreter (an interpreter's interpreter) that builds Druid's IR from the language interpreter. The meta-interpreter works as a message-directed translation, coping with Smalltalk's traditional *all computations are done through messages* (see Section 3.2). The unit of translation is a method defining a VM instruction (*i.e.,* bytecode and primitive instructions), and we perform aggressive inlinings. This is possible because the VM interpreter framework ensures by construction that all calls are monomorphic.

**Druid's mid end** (Figure 3, step (2)) At its core, Druid is a method-based compiler using an intermediate representation with a control flow graph in SSA form, maintaining use-def chains. Druid's mid-end optimizes its IR using traditional optimizations such as inlinings, constant folding, dead code elimination, and global value numbering [41].

**Druid's back end** (Figure 3, step (3)) Druid's target is Cogit [36], a dynamic machine code generator that uses abstract interpretation to implement peephole optimizations [47]. We kept interpreter changes to the minimum, limiting ourselves to behavior-preserving refactorings. Druid's back end first lowers the IR to accommodate it to the target, in our case, an x64-86-inspired register transfer language (see Section 4). Then, it stages computations that can be executed at JIT compile time (see Section 3.4). Finally, it generates source code that is deployed as a JIT compiler front end.

### 3.2 Intrinsics and Message-Directed IR Generation

Druid's front end generates the compiler intermediate representation using a message-directed translation, instead of a typical syntax-directed translation. That is, translation rules are based on messages and their selectors instead of syntactic elements. On the one hand, most messages are translated as normal calls. On the other hand, Druid specifies special translation rules for a subset of selectors called *intrinsics*.

**Definition 3** (Intrinsic)**.** *An intrinsic is a message-send that has a specific IR translation.*

The main reason for this decision is that the Pharo Virtual Machine is written in Pharo itself, a Smalltalk object-oriented dynamically-typed programming language, and follows the Smalltalk tradition: the language has a small syntax, and thus all operations including arithmetics *e.g.,* +, -, *, / and control flow are expressed as late-bound message sends. Thus, IR generation rules are written depending on the *message* found at call sites. This is possible because the VM is written in a style that assumes that all call sites have a single call target, allowing an efficient translation to C by the Slang VM generator [29, 37].

Our meta-compiler design further separates intrinsics into groups. Appendix A lists all our intrinsics by category. Guiding intrinsics that are important in our design are described separately in the sections that follow.

**Essential Intrinsics.** Intrinsics for standard Pharo code, implementing the semantics of standard library (*e.g.,* arithmetics) and control flow operations (*e.g., ifTrue:*).

**VM-specific Intrinsics.** Intrinsics for VM semantics, such as stack access (*e.g.*, push and pop operations), memory access, and type coercions. These have the same semantics as the Pharo-to-C translation done by Slang.

**Guiding Intrinsics.** Intrinsics to guide meta-compilation decisions. These intrinsics work mostly as interpreter no-ops except if specified differently.

Figure 4 shows an excerpt of code using sumSmallInteger:withSmallInteger:ifOverflow:, a VM-specific intrinsic that performs checked-arithmetics detecting overflow. Such an intrinsic is inlined when translating the interpreter to C, yielding no performance overhead, and is translated to a *jump if overflow* instruction in our JIT compiler backend.

```
Interpreter >> primitiveAdd
  ...
  result := self
    sumSmallInteger: smi1
    withSmallInteger: smi2
    ifOverflow: [ ^ self primitiveFail ].
  ...
```

```
JITFrontend >> gen_primitiveAdd
  ...
  self AddCq: -1 R: SendNumArgsReg.
  self AddR: ClassReg R: SendNumArgsReg.
  jump2 := self JumpOverflow: target.
  ...
```

**Figure 4** Meta-compilation of a primitive using an intrinsic performing SmallInteger addition with checked overflow.

### 3.3 Guiding Intrinsincs: Compiler and Interpreter Specific Code

The druidIgnore: and interpreterIgnore: intrinsics allow fine-tuning the compilation scope to ignore expensive compilation cases or to rewrite parts of the code that must be performed differently in the interpreter or the compiler. They allow us to control code that will be ignored by the JIT compiler or the interpreter, respectively. Our most common usage for such intrinsics is omitting compilation of interpreter-specific optimizations such as static type predictions [27], which are subsumed by polymorphic inline caches in the JIT compiled code [28].

Figure 5 illustrates the usage of druidIgnore: to meta-compile closure activation. During interpretation, a static-type prediction avoids expensive lookups. In JIT-compiled code, such optimization becomes useless in the presence of PICs.

### 3.4 Druid Staging Pass

Druid implements a staging pass that runs at the end of the pipeline, after all optimizations but just before register allocation and code generation. The staging pass is designed as a forward sparse data-flow problem following SSA use-def chains, similar to a constant propagation analysis [1], identifying JIT-constant expressions. After the analysis, all expressions identified as JIT-constant are marked for staging during the code generation pass.

```
Interpreter >> bytecodePrimValue
  ...
  self druidIgnore: [
    | rcvr isBlock |
    rcvr := self stackTop.
    isBlock := self isInstanceOfClassBlockClosure: rcvr.
    isBlock ifTrue: [ ^ self primitiveFullClosureValue ] ].

  self normalSendSpecialSelector: 25 argumentCount: 0
```

**Figure 5** Excerpt of the interpreter using the druidIgnore: intrinsic to avoid the compilation of static type prediction during closure application. This optimization is subsumed by message-send optimizations such as polymorphic inline caches.

To support expressions other than constants for staging, we extended the interpreter with intrinsic functions to explicitly annotate expressions for staging, which is equivalent to a *lift* operation in other meta-compilers [44, 25]. Figure 6 presents an excerpt of code that illustrates how we have used this intrinsic to stage, for example, the load of method literals, which remain constant during the life-cycle of a method.

```
Interpreter >> literal: offset ofMethod: methodPointer

  ^ self druidStageable: [
     objectMemory
       fetchPointer: offset + LiteralStart
       ofObject: methodPointer ]
```

**Figure 6** Excerpt of the interpreter using the druidStageable: intrinsic to lift literal load operations as compile time constants.

### 3.5 Compilation Exit Points

*Compilation exit points* are annotations (*pragmas* in Pharo parlance [19]) marking interpreter methods that must avoid JIT compilation. Instead, those methods must output code performing a runtime call. Within bytecode instructions, exit points deoptimize the current stack frame spilling all registers into the stack, and perform a runtime call through a trampoline. Within primitive instructions, exit points skip to the end of the primitive instruction, compiling a fall-through so the compiler framework generates a call to the interpreter.

*Compilation exit points* are useful to scope the work of the JIT compiler, and the usage in our experiments was two-fold. First, from a JIT compiler design standpoint, it allows us to leave slow paths outside of the JIT compiler scope. Second, from an iterative development standpoint, it allowed us to incrementally extend the JIT compiler and have early, yet incomplete, working versions to experiment.

Figure 7 shows the annotation druidExitPoint that indicates to Druid to ignore the compilation on the method contextSize and all paths calling it.

```
Interpreter >> contextSize
  "Special version of primitiveSize for accessing contexts."
  <druidExitPoint>
  ....
```

**Figure 7** The exitPoint annotation allows Druid to skip complex operations (*e.g.,* access to context information) that cause long compilation times, ignoring them by the JIT compiler and deoptimizing to the interpreter instead.

### 3.6 Compilation Metadata

We added metadata as method annotations to each instruction. Metadata declares the number of arguments of primitive functions, the delta left in the stack [14, 22], if the instruction defines a suspension point, and if the instruction requires a stack-frame for its execution. This metadata is required mainly by the target JIT compiler backend to decide frame-ful/less compilation and machine-code mappings for deoptimization. Moreover, our meta-compiler uses this metadata to transform the stack-based calling convention in the interpreter to the register-based calling convention in the target JIT compiler. Appendix A lists all druid annotations. We plan in the future to extract such annotations automatically using static analysis.

Figure 8 shows two annotated methods defining respectively a primitive and a bytecode instruction in the Pharo VM. The first method, defining the primitive that adds two small integers, declares the number of arguments to properly map the stack calling convention to the register-based one. The second method, defining the fast-path add bytecode, declares that the instruction contains a message-send and that it's program counter should be mapped because it contains a deoptimization point.

```
Interpreter >> primitiveAdd
  <numberOfArguments: 1>
  ....
```

```
Interpreter >> bytecodePrimAdd
  <compilationInfo: #isMapped>
  <druidInfo: #hasSend>
  ....
```

**Figure 8** Metadata of language instructions as annotations in the interpreter definitions.

## 4 Targeting an Existing Compiler Framework: Cogit

For completeness, this section presents several implementation details that were important during the development of our prototype. Although important, we believe many of these points are implementation-specific and will be relevant for similar implementations only.

### 4.1 Targeting the Cogit Dynamic Machine Code Generator

Druid compilation targets Cogit, a JIT compiler backend with support for different architectures [36], developed originally for the OpenSmalltalk-VM [37, 29]. Cogit is a method-based dynamic machine code generator that takes as input a Pharo method. Cogit frontends must define an IR generator for each bytecode/primitive instruction. Our solution meta-compiles each bytecode and primitive handler defined in the interpreter to generate the IR generators for Cogit, as illustrated in Figure 3.

IR generators are defined using a register-transfer language, namely CogRTL, to build its intermediate representation (IR) from the source bytecode. CogRTL is designed as an intel-inspired 2 address code IR (2AC). Cogit uses a fixed list of registers with virtual names (*e.g.,* `TempReg`, `ReceiverReg`, `ClassReg`) that are manually mapped ahead of time to machine registers by JIT-compiler developers.

**Cogit register allocation** Primitive method IR generators are defined as code templates in CogRTL. In this case, Cogit works mostly as an assembler: JIT frontend developers implement the instruction entirely in CogRTL and perform register allocation manually. Instead, bytecode compilation uses an abstract interpreter to optimize across the boundaries of each instruction [47]. The main optimization done by this abstract interpreter is register allocation as a peephole optimization: registers are tracked on an abstract stack, assigned dynamically, and spilled when needed. The current register allocator uses only the value stack, reducing the scope of register allocation to single statements.

**Cogit runtime interactions** As explained above, Cogit transforms stack-based bytecode to register-based machine code and tries to avoid stack accesses across bytecode boundaries. Cross-runtime calls are done through trampolines, small machine-code routines that implement the system calling convention when jumping from generated machine code to runtime, and vice-versa. Register values must be spilled to the stack when performing machine code to runtime calls. This is because runtime functions assume that all data is on the stack, *e.g.,* the GC traverses the stack to find graph roots. Stack spilling needs to be manually managed by the JIT front end developer and is not automatically performed by trampolines.

**Cogit polymorphic inline caches** The Cogit JIT compiler infrastructure has built-in support for dynamically-bound message-sends using (polymorphic) inline caches (ICs) [28]. ICs are implemented by machine code patching. They support unbound call sites linked to a *miss* trampoline, monomorphic call sites linked directly to methods, and polymorphic and megamorphic call sites linked to machine code stubs.

### 4.2 Accommodating Druid to the Cogit JIT Compiler

As mentioned in Section 3, the Druid compilation scope is one interpreter handler. From this, it generates a control flow graph intermediate representation (IR) in 3 address code (3AC) SSA form [16]. After applying all optimizations, Druid lowers the IR to accommodate CogRTL.

**Accommodating to Cogit register allocation** Druid performs a register allocation pass that performs 3AC to 2AC translation and targets Cogit's registers using the constraint approach found in [38]. When meta-compiling primitive instructions, Druid's register allocator assigns Cogit's fixed virtual registers. When meta-compiling bytecode instructions, we output instructions that will dynamically –at JIT compile time– allocate registers following the values in the abstract stack. Examples of both meta-compiling instructions are present in Appendix B.

**Accommodating to Cogit runtime interactions** When Druid translates runtime interactions, it must take care of spilling registers to the stack and respect the trampoline calling convention. Our current implementation supports only pre-existing trampolines.

**Accommodating to Cogit polymorphic inline caches** We decided in our current implementation to reuse existing support of ICs and call site linking. Message-sends are expressed in the interpreter through intrinsic functions generating a `DRSend` instruction. Druid translates `DRSend` instructions during code generation into a patchable call-site RTL sequence.

### 4.3 Changes to the Compiler Architecture

Besides introducing intrinsic functions and annotations in the interpreter, we performed several changes to the Cogit architecture. First, the existing frontend was hand-tailored to compiling Pharo code and thus it implemented workarounds to support advanced features instead of exposing well-defined APIs. Moreover, the compiler architecture was originally engineered to compile the entire language, preventing us from incrementally developing the compiler and experimenting with intermediate results.

**Supporting incremental compiler development** We extended the runtime to avoid the unnecessary re-compilation of methods that previously failed compilation. The Cogit JIT compiler originally worked as a *compile all or nothing* compiler, failing compilation and falling back to the interpreter as soon as a compilation error happened *e.g.,* because register allocation did not satisfy some particular constraints, or a bytecode did not have a defined IR generator. This architecture worked well for the hand-written frontends where all the language was specified. However, this incurred a large JIT compilation overhead for our first experiments when working with partial translations.

**Partial compilation support** Following the previous point, we also extended the compiler with support for arbitrary *deoptimization points*. Deoptimization points allowed us to focus on fast execution paths and jump to the interpreter in the case of less common slow paths. Deoptimization points spill the value of registers on the stack, transfer the control to the runtime, deoptimize the stack frame, and transfer control to the interpreter.

**Improving the target compiler API** We extended the compiler API, particularly the abstract stack API, to simplify meta-compilation. The existing front end was hand-tailored and depended on undocumented workarounds on the compiler backend. Moreover, it used non-homogeneous calling conventions when activating methods, block closures, and even different trampolines. Instead of pushing such complexity to our meta-compiler, we improved the exposed APIs and simplified calling conventions.

# 5 Evaluation

To answer the research questions in Section 1, we evaluate how our prototype performs in comparison with the interpreter and the hand-written baseline JIT compiler of the Pharo VM.

Both JIT compilers—one using the Druid-generated frontend and the other using the hand-written version—use Cogit as the backend, thus they just diverge in the IR generators in the frontend.

## 5.1 Setup and Methodology

We built a set of four different variations of the Pharo VM. Each variation has a different behavior regarding interpretation vs. compilation, the amount of code that is compiled and supported optimizations:

***InterpreterOnly.*** Pharo VM without JIT compiler, only an interpreter implementing indirect threading. It implements the entire language encompassing 240 bytecodes and 263 primitive instructions. It is used as the baseline for speed-up measurements.

***FullManualJITVM.*** Pharo VM extending *InterpreterOnly* with a handwritten compiler frontend, in production for at least 10 years. It supports the compilation of 240 bytecodes and 130 primitive instructions. This production VM includes other optimizations such as *static type prediction* and *dynamic super-instructions*.

***DruidJITVM.*** Pharo VM extending *InterpreterOnly* with a Druid autogenerated compiler front end. It supports the compilation of 236 bytecodes and 52 primitive instructions. The four bytecodes missing from the above implement unused/experimental instructions from the Sista bytecode set [8] and the stack frame reification instruction (*i.e.,* `pushThisContext`) that we considered part of the *slow paths* of a program (*e.g.,* exceptions, debugging).

***MirrorJITVM.*** Pharo VM using a subset of *FullManualJITVM* compiling exclusively the same bytecodes and primitive instructions as *DruidJITVM*. We use this subset to fairly compare our solution's quality against the handwritten version in *FullManualJITVM*.

We run a set of classical benchmark programs written in Pharo on each VM. We use the Rebench framework [34, 35] to run the benchmarks. We run all benchmarks using 30 VM iterations. Micro benchmarks use 10 in-process with two (ignored) warmup iterations. This number of iterations showed enough to measure the behavior of the non-optimizing baseline JIT compilers. Macro benchmarks, consisting of running the test suite of several packages, use a single in-process iteration. For all the results below, we report the average of the execution time of each in-process iterations.

We build each VM and run the benchmarks using a Linux Debian 5.10.140-1 distribution on a AMD64 CPU E5620 @ 2.40GHz processor with 16 GB DDR3 of RAM.

### 5.2 Benchmarks

**Table 1** Relative speed-up between the VMs per benchmark. Higher is better. (D = *DruidJITVM*; I=*InterpreterOnly*; FM=*FullManualJITVM*; M=*MirrorJITVM*)

| Speed-up | | Optimized | | | Unoptimized | | |
|---|---|---|---|---|---|---|---|
| **Benchmark** | D/I | D/FM | D/M | M/FM | D/FM | D/M | M/FM |
| *BinaryTrees* | 3.08 | 0.66 | 0.66 | 1.00 | 0.69 | 0.69 | 1.00 |
| *Chameleons* | 1.50 | 0.95 | 0.97 | 0.98 | 0.96 | 0.97 | 0.99 |
| *ChameneosRedux* | 2.96 | 0.80 | 0.78 | 1.03 | 0.82 | 0.84 | 0.98 |
| *DeltaBlue* | 2.90 | 0.64 | 0.65 | 0.97 | 0.44 | 0.87 | 0.51 |
| *Fasta* | 0.92 | 0.39 | 0.65 | 0.60 | 0.45 | 0.69 | 0.65 |
| *KNucleotide* | 2.20 | 0.89 | 0.93 | 0.95 | 0.79 | 0.96 | 0.82 |
| *Mandelbrot* | 2.08 | 0.56 | 0.56 | 0.99 | 0.60 | 0.59 | 1.01 |
| *Meteor* | 2.32 | 0.51 | 0.52 | 0.98 | 0.67 | 0.69 | 0.97 |
| *NBody* | 2.28 | 0.61 | 0.63 | 0.97 | 0.62 | 0.69 | 0.89 |
| *PiDigits* | 1.08 | 0.97 | 0.98 | 0.99 | 0.99 | 0.98 | 1.01 |
| *RegexDNA* | 2.86 | 0.59 | 0.81 | 0.72 | 0.45 | 0.86 | 0.52 |
| *ReverseComplement* | 1.68 | 0.88 | 0.70 | 1.26 | 0.75 | 0.72 | 1.04 |
| *Richards* | 5.09 | 0.69 | 0.72 | 0.96 | 0.52 | 0.78 | 0.67 |
| *Slopstone* | 1.27 | 0.42 | 0.78 | 0.54 | 0.54 | 0.89 | 0.61 |
| *SpectralNorm* | 1.42 | 0.61 | 0.60 | 1.01 | 0.64 | 0.64 | 1.00 |
| *ThreadRing* | 0.95 | 0.91 | 0.98 | 0.93 | 0.99 | 1.05 | 0.94 |
| *Compiler* | 1.52 | 0.89 | 0.89 | 1.00 | 0.89 | 0.93 | 0.96 |
| *Microdown* | 1.09 | 0.59 | 0.61 | 0.98 | 0.68 | 0.72 | 0.95 |
| *Network* | 1.29 | 0.80 | 0.81 | 0.99 | 0.84 | 0.94 | 0.90 |
| *Startup* | 1.40 | 0.74 | 0.76 | 0.98 | 0.81 | 0.89 | 0.91 |
| **Best** | 5.09 | 0.97 | 0.98 | 1.26 | 0.99 | 1.05 | 1.04 |
| **Worst** | 0.92 | 0.39 | 0.52 | 0.54 | 0.44 | 0.59 | 0.51 |
| **Average** | 1.99 | 0.70 | 0.75 | 0.94 | 0.71 | 0.82 | 0.87 |

Our set of classical **micro benchmarks**[2] is as follows:

***BinaryTrees.*** An adaptation of *Hans Boehm's GCBench* for Garbage Collection.
***Chameleons.*** Creates many threads to wait for mutex semaphores.
***ChameneosRedux.*** Small problem size for *Chameleons* benchmark.
***DeltaBlue.*** Classic object-oriented *constraint solver*.
***Fasta.*** DNA sequence generation algorithm based on weighted random selection.
***KNucleotide.*** Map the DNA letters into a hash table to accumulate count values.
***Mandelbrot.*** Plot the Mandelbrot set on an N-by-N bitmap.
***Meteor.*** Uses *ByteStrings* to solve a puzzle.
***NBody.*** Classic n-body simulation of the solar system.
***PiDigits.*** Generates and prints the first N digits of Pi.
***RegexDNA.*** A simple regex pattern and actions to manipulate FASTA format data.
***ReverseComplement.*** Computes the reverse complement of a DNA sequence.
***Richards.*** Simulates a *task dispatcher* on a multitasking operating system.
***Slopstones.*** *Smalltalk Low-level OPeration Stones*, a series of low-level operations.
***SpectralNorm.*** Calculate the spectral norm of a N-by-N matrix.
***ThreadRing.*** A token ring algorithm using threads and mutex semaphores.

In addition, we run some applications as **macro benchmarks**:

***Compiler.*** Tests for source-to-bytecode Pharo compiler.
***Microdown.*** Tests for markup documentation library.
***Network.*** Tests for network communication library.
***Startup.*** Just evaluate `1 + 1`, measure the time for starting up the system.

### 5.3 RQ1: Does our solution generates JIT compilers that improve the performance of the interpreter?

This research question evaluates whether a Druid autogenerated baseline-JIT compiler frontend performs better than a pure interpreter. Figure 9 shows, for each benchmark, the execution time relative to *InterpreterOnly*. *InterpreterOnly* remains in 1× as it is our baseline. These same results are shown in the first column of Table 1.

Our results show that the generated JIT compiler frontend presents an average improvement of 2×. In the best scenario, it shows improvements up to 5.09× (*Richards*). From our macro benchmarks, we see a speedup between 1.09× (*Microdown*) and 1.52× (*Compiler*).

Only three negative results can be observed in the table. First, we can also observe that *PiDigits* and *ThreadRing* are not very sensitive to any JIT compilation, showing similar improvements for all compilers. Moreover, we observe that our solution negatively impacts the performance of *Fasta*, by 8 %, when compared to the interpreter. From the results in the following Section 5.4, we see that the results of *Fasta* are

[2] Implementations are in https://github.com/guillep/SMark visited on 2024-02-04.

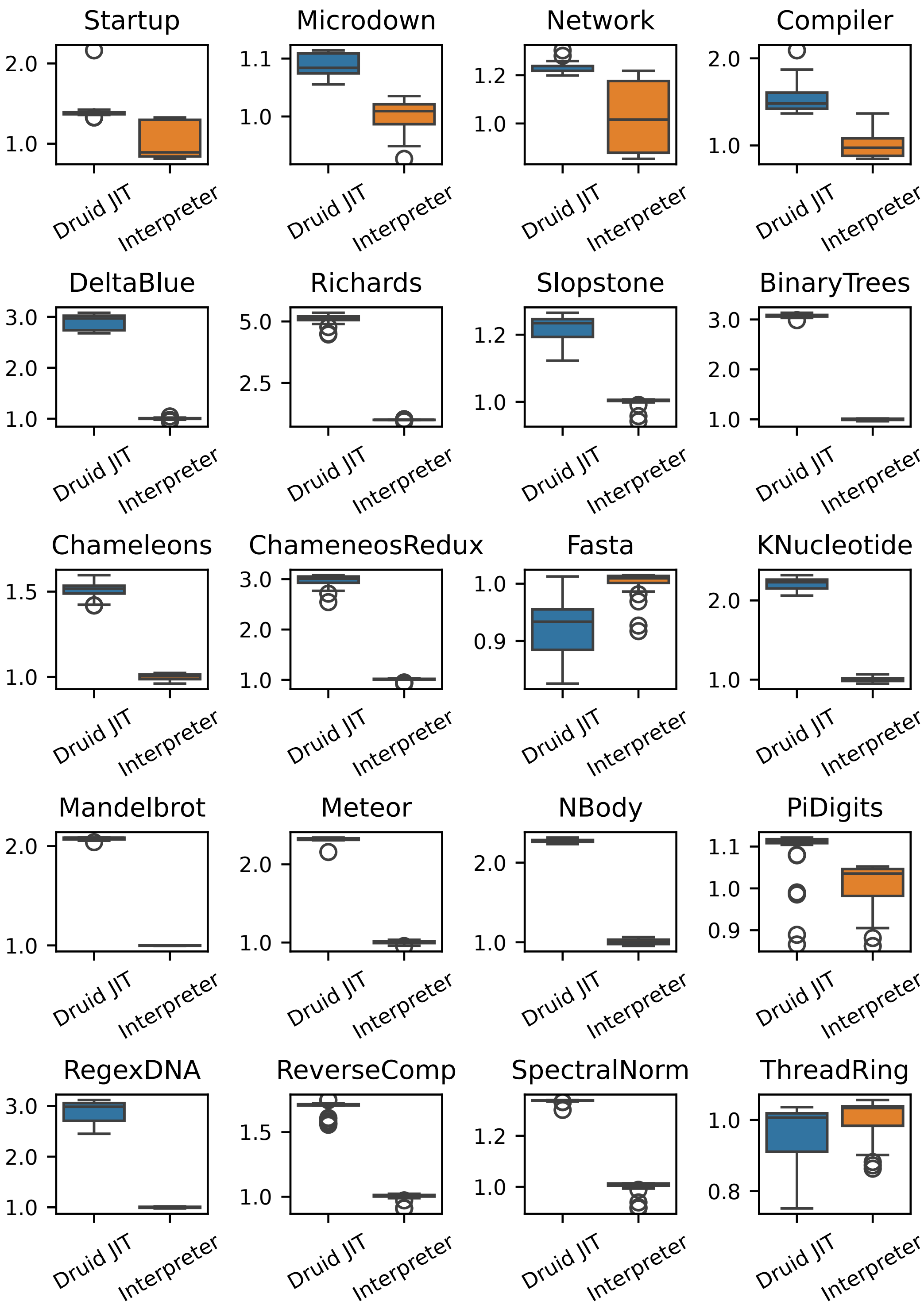

■ **Figure 9** Speed up of the Pharo VM using the JIT compiler generated automatically by our approach (*DruidJITVM*) compared to *InterpreterOnly* (baseline). Higher is better.

probably caused by low-quality generated machine code in a common case of that benchmark.

> **RQ1 Conclusion.** Our solution generates JIT compilers that improve the performance of just interpreted VMs by an **average of** 2×**, up to** 5× **faster**.

### 5.4 RQ2: Is the JIT compiler generated by Druid competitive in performance against the handwritten version?

This research question evaluates how well Druid performs in comparison with a 10-year-old matured JIT compiler front end. For this purpose, we compare *DruidJITVM* with *FullManualJITVM* and *MirrorJITVM*. We use *FullManualJITVM* as an upper bound and main goal: our objective is that *DruidJITVM* approaches the performance of *FullManualJITVM*. However, such a comparison does not yield fair results as *FullManualJITVM* translates more code than our prototype.

To better understand the quality of our generated code against the handwritten version, we compare *DruidJITVM* to *MirrorJITVM*. The translations of *MirrorJITVM* and *DruidJITVM* define an isomorphism: all translations defined in *DruidJITVM* are defined in *MirrorJITVM* and vice-versa. However, the translations of *MirrorJITVM* were extracted from *FullManualJITVM*.

Figure 10 shows a boxplot of our benchmarks, with all results plotted relative to the *InterpreterOnly*. Table 1 (Optimized) shows the ratio of improvement between all three solutions. Results should be interpreted as follows:

- When *DruidJITVM* and *MirrorJITVM* are close, our generated JIT compiler is as good as the baseline. Conversely, if they are distant, this indicates potential improvements in generated front end.
- When *MirrorJITVM* and *FullManualJITVM* are close, the set of translations (bytecodes and primitives) properly represents the benchmark working set. Conversely, if they are distant, this indicates missing translations in our prototype.

From our results, *DruidJITVM* achieves on average 0.70× the performance of *FullManualJITVM*, and particularly 0.59× (*Microdown*) and 0.89× (*Compiler*) for our macro-benchmarks. *MirrorJITVM*, on the other side, is almost as fast as the handwritten one, 0.94× on average, except for benchmarks *Slopstone*, *Fasta* and *RegexDNA* where it is slower. *DruidJITVM* achieves on average 0.75× the performance of *MirrorJITVM*.

Using these results, we conclude that *DruidJITVM* supports a good set of language instructions (the same as the *MirrorJITVM*), but the quality of the JIT compiler produced by our meta-compiler is not yet as good as the one found in *FullManualJITVM* that has been matured for 10 years. For example, optimizations missing in *DruidJITVM* that are present in *FullManualJITVM* are:

1. *static type prediction* for arithmetics and bit-wise manipulation [27].
2. *dynamic super-instructions* for comparisons and jumps [15].

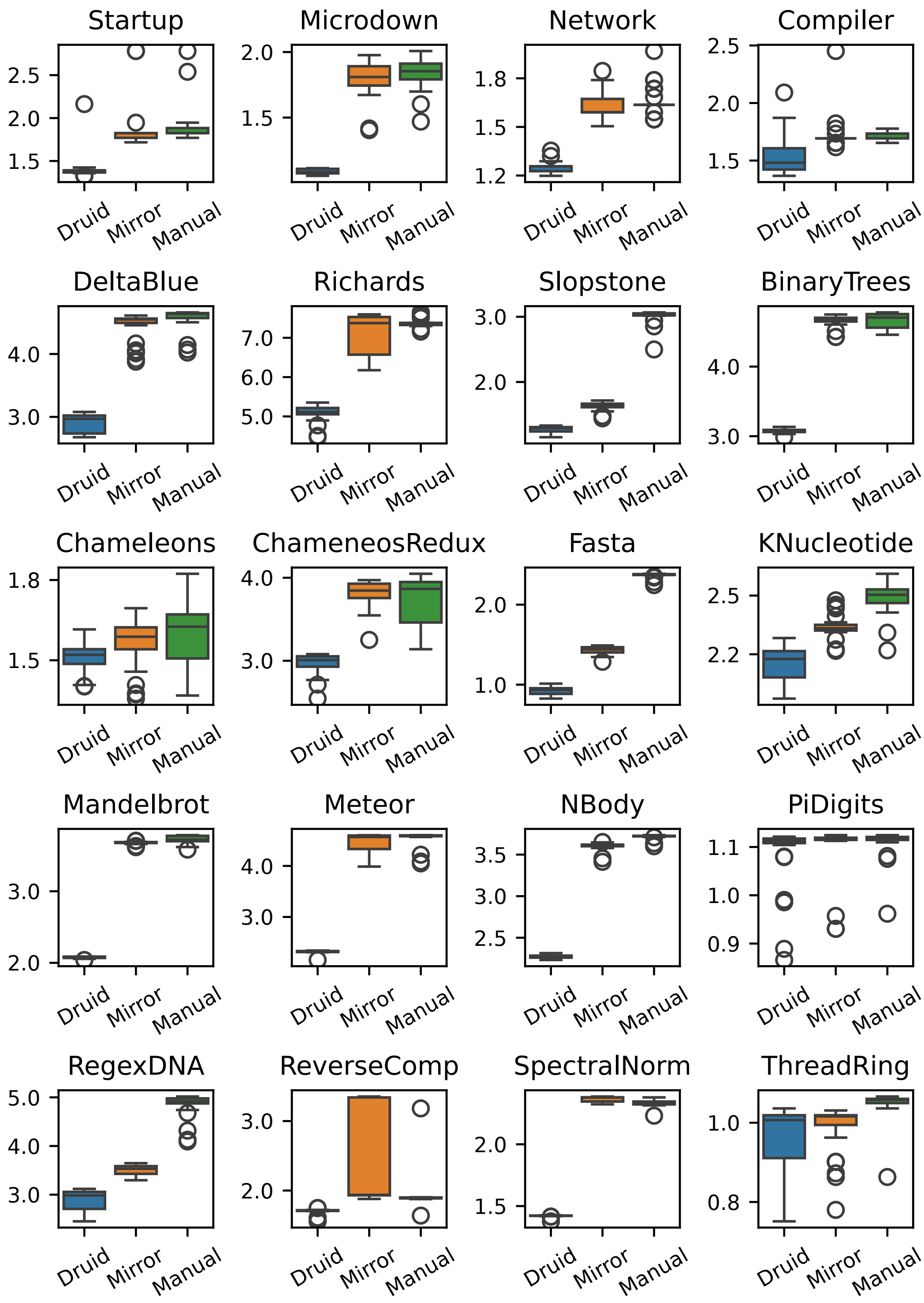

**Figure 10** Speed up of VM using different versions of the JIT compiler: *DruidJITVM*, *MirrorJITVM* and *FullManualJITVM*. *InterpreterOnly* is used as the baseline. Higher is better.

To measure the impact of those optimizations in our benchmarks, we re-compiled all of them using unoptimized bytecode (avoiding the optimization cases in the VM). The results of re-run the benchmarks using the unoptimized bytecode are present in Table 1 (Unoptimized). We compare the results between optimized and unoptimized in Figure 11.

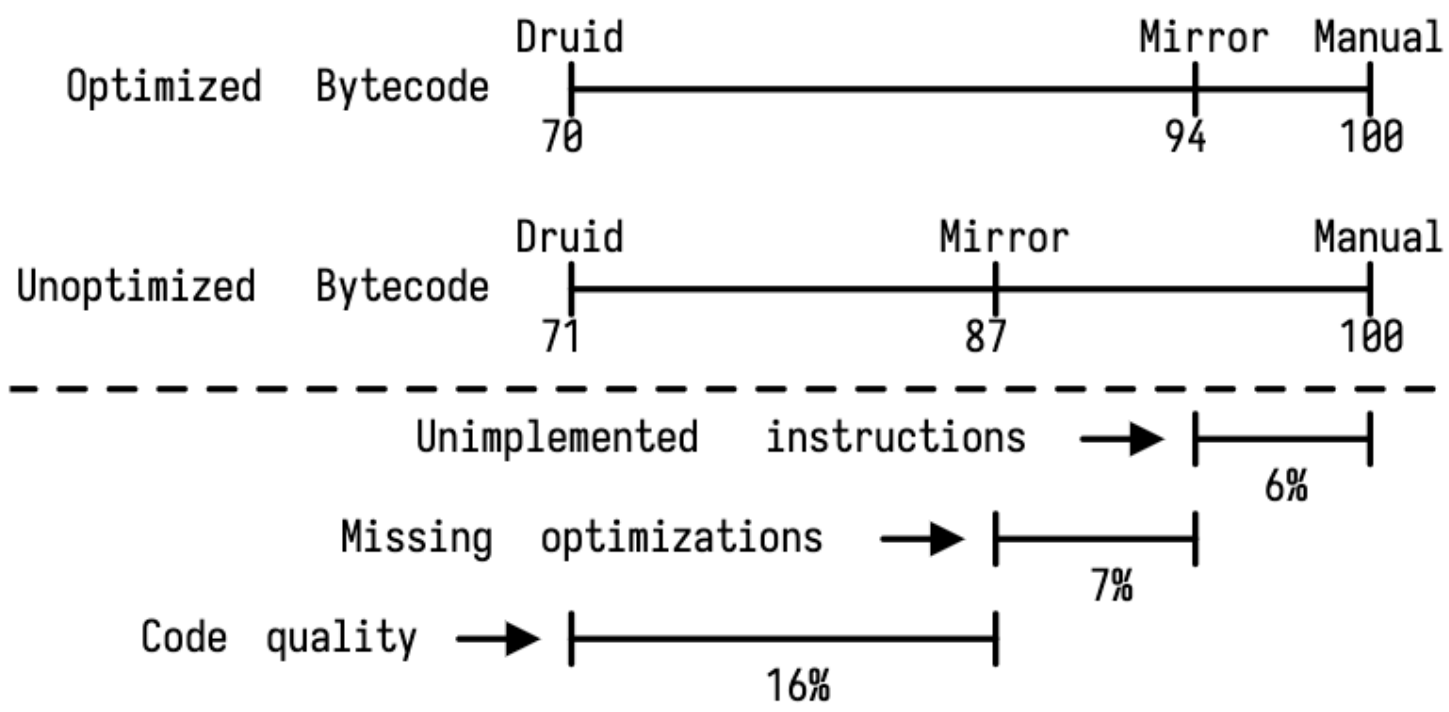

**Figure 11** Estimating the impact of generated code quality, missing optimizations, and unimplemented instructions between *DruidJITVM* and *FullManualJITVM*.

The experiments using the unoptimized bytecode suggest that the 30 % of the performance gap between *DruidJITVM* and *FullManualJITVM* is composed by:

- 6 % of unimplemented VM instructions in the translation set.
- 7 % of missing optimizations (*i.e., static type prediction* and *dynamic super-instructions*).
- The remaining 16 % comes from the quality of the JIT-compiled code.

> **RQ2 Conclusion.** Our solution generates a JIT compiler that **achieves the** 0.70× **of the speed** of the handwritten version matured for 10 years. Improving the set of translated instructions shows a potential performance improvement of 6 %. Other 7 % comes from missing optimizations in our meta-compiler such as static type predictions. The rest of the difference is due to the quality of the JIT-compiled code.

### 5.5 RQ3: Does our generated approach increase the overhead at compile time?

To assess that our approach does not generate JIT compiler front ends that introduce significant compilation overhead, we measure the compilation time per benchmark and compute what percentage of the total execution time it represents. Then we compare those values with the obtained for the *FullManualJITVM*. Results are presented in Table 2.

On average, the JIT compile time represents ∼6 % of the total run time for both VMs. There are no significant differences between *DruidJITVM* and *FullManualJITVM*. The biggest difference between both VMs is, for the micro-benchmarks, *PiDigits* and *ThreadRing*, already explained in Section 5.3. Moreover, the *Microdown* macro-benchmark presents a compilation overhead 2.3× bigger than *FullManualJITVM*. The authors believe that it represents a *compilation pathological case* in *DruidJITVM* due to the big JIT compilation overhead presented in this benchmark in comparison with the others. Figure 12 shows the absolute execution and compilation time for all benchmarks running the *DruidJITVM*.

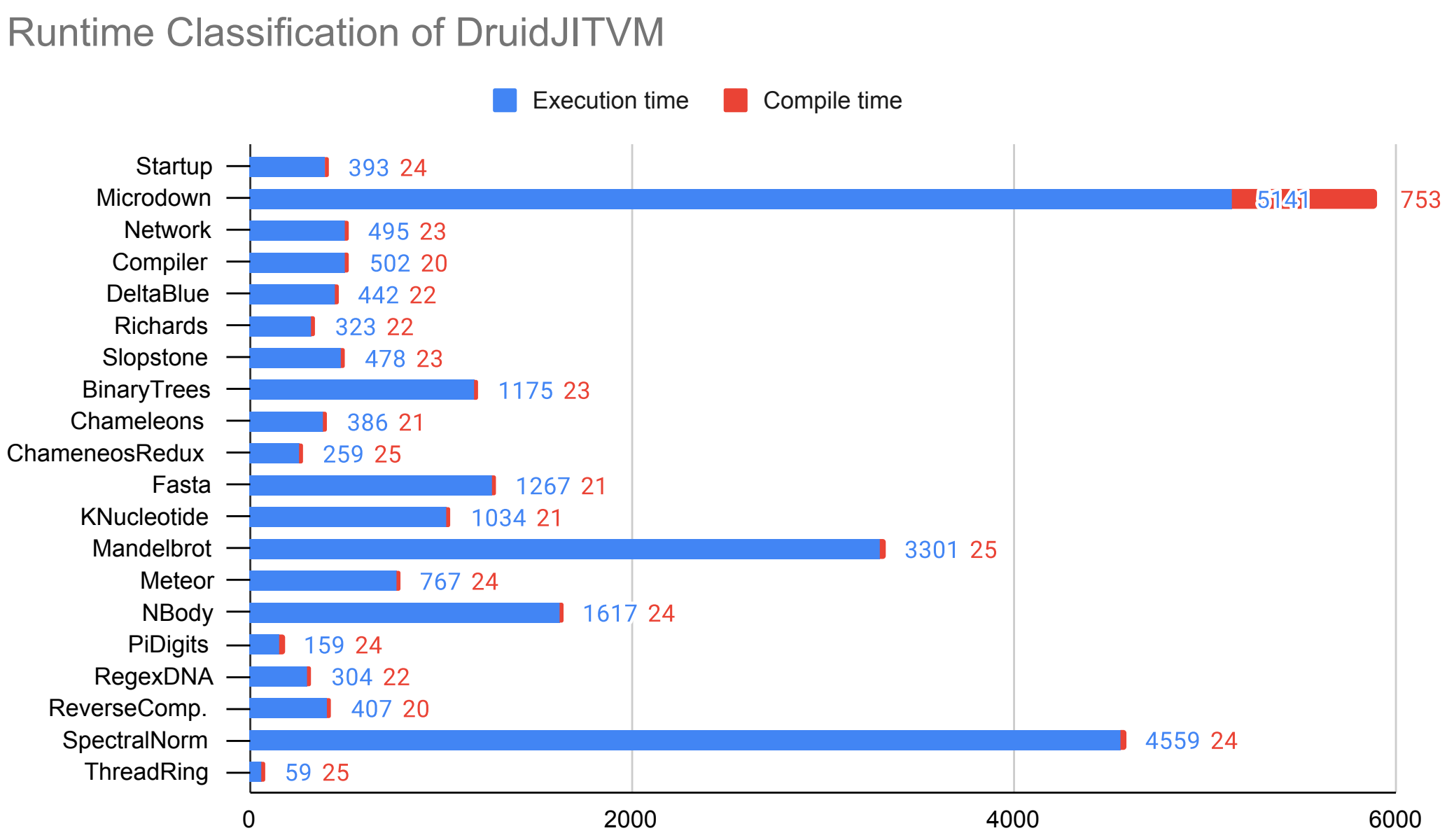

**Figure 12** Compile and execution times for *DruidJITVM*.

Our benchmark results show, on average, a tiny overhead increment of 0.18 % between *DruidJITVM* and *FullManualJITVM*. This means that our solution does not increase the overhead at JIT compile time.

> **RQ3 Conclusion.** Druid's generated JIT compiler adds only 0.18 % more overhead, on average, compared to the manually developed one, at JIT compile time. Both represent nearly the same portion of the total execution time for most benchmarks. This means that **our solution does not introduce significant additional overhead to the production VM at JIT compile time**.

### 5.6 RQ4: How much development effort is affected by our automatic approach?

In addition to the achieved performance and relative speed regarding a handwritten JIT compiler, we are interested in evaluating the impact of our solution on the development effort. So far, language developers have required compiler knowledge to extend the Pharo Virtual Machine.

**Table 2** Compile time relative to total runtime. The last column shows the difference between VMs. Lower is better.

| **Compile Time relative to total run time** | | | |
|---|---|---|---|
| **Benchmark** | *DruidJITVM* | *FullManualJITVM* | **difference** |
| *BinaryTrees* | 1.92 % | 2.50 % | -0.58 |
| *Chameleons* | 5.22 % | 4.07 % | +1.15 |
| *ChameneosRedux* | 8.96 % | 7.06 % | +1.90 |
| *DeltaBlue* | 4.72 % | 5.41 % | -0.69 |
| *Fasta* | 1.59 % | 2.67 % | -1.08 |
| *KNucleotide* | 1.99 % | 1.99 % | 0.00 |
| *Mandelbrot* | 0.75 % | 0.94 % | -0.19 |
| *Meteor* | 2.97 % | 5.08 % | -2.11 |
| *NBody* | 1.46 % | 1.57 % | -0.11 |
| *PiDigits* | 13.33 % | 8.78 % | +4.55 |
| *RegexDNA* | 6.75 % | 9.56 % | -2.81 |
| *ReverseComplement* | 4.73 % | 7.17 % | -2.44 |
| *Richards* | 6.24 % | 5.77 % | +0.47 |
| *Slopstone* | 4.64 % | 8.56 % | -3.92 |
| *SpectralNorm* | 0.52 % | 0.60 % | -0.08 |
| *ThreadRing* | 29.44 % | 25.34 % | +4.10 |
| *Compiler* | 3.83 % | 4.48 % | -0.65 |
| *Microdown* | 12.77 % | 5.57 % | +7.20 |
| *Network* | 4.35 % | 4.57 % | -0.22 |
| *Startup* | 5.86 % | 6.63 % | -0.77 |
| **Best** | 4.64 % | 8.56 % | -3.92 |
| **Worst** | 12.77 % | 5.57 % | +7.20 |
| **Average** | 6.10 % | 5.92 % | +0.18 |

This section estimates the development efforts by considering the amount of generated code (in KLOCs), the amount of handwritten code (in KLOCs) in the existing front end, and the number of modified interpreter call sites. We measured lines of code by selecting the involved Pharo methods and counting the lines of the source code, including comments and white lines. Figure 13 show our measurements for the different components, with and without applying Druid. Druid being a general compiler infrastructure, we considered the common parts (the IR, the optimizations) separately from the meta-compilation specific parts. We considered:

**Handwritten compiler frontend.** All IR generator methods (with prefix `gen` in `StackToRegisterMappingCogit` hierarchy, for bytecodes, and `CogObjectRepresentationFor64BitSpur` hierarchy, for primitives; except the ones in `Cogit` superclass, they are part of the backend).

**Generated frontend.** All methods generated by the meta-compiler (with prefix gen_ in DruidJIT class).

**Druid compiler infrastructure.** All methods in the Druid package, excluding tests, experimental features, and code inside the Druid-Cogit subpackage.

**Druid cogit extensions.** All methods in the Druid-Cogit and Druid-Staging subpackage.

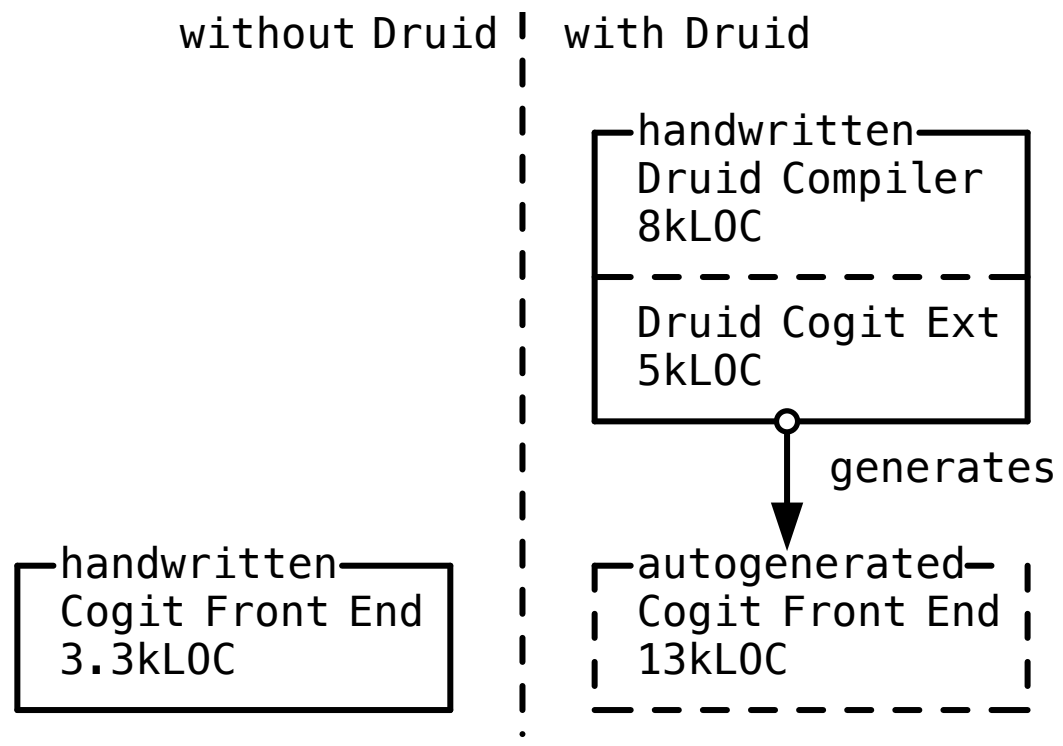

**Figure 13** Lines of code for the different parts of our solution. Druid is a handwritten compiler comprising 8KLOCS of infrastructure plus 5KLOC of Cogit extension. It generates 13KLOC for the JIT compiler frontend, equivalent to the handwritten 3.3KLOC in the production VM.

In each of the discussions below, we put in perspective such numbers, as manually written code tends to contain abstractions and avoid code repetitions for maintenance *e.g.,* common code extracted in methods.

**Separation of concerns** The JIT compiler is one of the most complex components to maintain inside a virtual machine [31], by avoiding handwriting it we are lowering the barrier for many developers, non-experts in VMs, who want to contribute to the language. In the context of an open-source project, as in the case of Pharo, the *democratization* of the VM code is an important goal. Thanks to Druid's architecture, language developers concentrate on an annotated interpreter and do not need compiler expertise to obtain a JIT compiler, thus reducing the maintenance of a specific language implementation. In our experiments, the Druid-generated JIT front end encompasses a total of 13.4KLOC equivalent to 3.3KLOC handwritten and maintained by VM-compiler developers over the last 10 years. Such a difference between our generated frontend and the manually maintained (~4×) is generated by Druid's aggressive inlinings.

**Interpreter and runtime maintenance** Our solution has minimal impact on language developers who need only to use intrinsics and annotations to guide meta-compilation. In our experiments, we did our best to minimize modifications to the interpreter, ensuring it remained backward compatible. To evaluate the cost of interpreter modification, we collected all interpreter call sites using our intrinsics (described in Section 3). Such interpreter modifications represent only **60 modified call sites**. However, it is important to notice that experience in language implementation, profiling, and optimization is required to achieve decent performance.

**Trading-off meta-compiler maintenance** The main trade-off of our architecture is that language maintenance comes at the cost of higher VM infrastructure maintenance. However, Druid is a compiler infrastructure with a modular architecture comprising 13KLOC of Pharo code. 8KLOC out of 13KLOC make up Druid's common compilation infrastructure: IR instructions and optimization passes. This subset of Druid works as a standard optimizing compiler only. The additional 5KLOC implement passes required for meta-compilation: meta-interpretation of VM semantics, intrinsic implementations, and code generation targeting Cogit. These passes, used for JIT-compiler frontend generation, are in the same order of magnitude as the manually maintained front end.

This trade-off is softened because of many reasons. First, the maintenance of Druid and the Pharo language can be assured by developers with different skill sets. Moreover, the Pharo VM code, written in Slang [37], limits the libraries, idioms, and abstractions that can be used in VM code. However, Druid code is not limited by those constraints: developers can write idiomatic Pharo code. Second, compiler generation can easily accommodate tasks such as helping evolve an existing compiler *e.g.,* by evolving the compiler IR and autogenerating the front ends targetting it. Finally, as we said above, Druid implements a traditional multi-pass compiler infrastructure, which can be leveraged in other contexts.

**RQ4 Conclusion.** Druid is an optimizing meta-compiler comprising 13K lines of pure Pharo code. It generates a total of 13.4KLOC of JIT-compiler front end, equivalent to handwritten 3.3KLOC. Modifying and maintaining those lines of code requires a completely different skill set than maintaining a language interpreter, which only required changes in 60 call sites. The size in KLOC of custom extensions are in the same order of magnitude as the manually written frontend, while separating the concerns of language semantics and its compilation. Moreover, Druid is implemented as a traditional multi-pass compiler infrastructure, which can be leveraged in other contexts. All in all, we believe that such an architecture turns language VMs **easier to maintain and evolve in the long run**.

### 5.7 Discussion and Lessons Learned

This work presents the viability of meta-compilation in translating interpreters defined by instruction handlers into baseline JIT compiler generators using a traditional multi-pass compiler enhanced with specialized frontend and backend extensions. Targeting an existing JIT compiler architecture allow us to focus on the meta-compilation of the language's features, avoiding JIT compilation details like specific architecture instructions encoding.

Our approach confirms that meta-compilation can be achieved as an ahead-of-time solution, without being present during runtime execution and enabling development in a language different from the host VM. It also presents the necessity of including specific annotations or intrinsics within the interpreter code to guide meta-compilation effectively, as well as similar approaches.

While the performance achieved is a significant improvement over interpreters, the results also show space for further optimizations and quality improvements of the jitted code to close the gap with handwritten JIT compiler performance.

## 6 Related Work

In the last two decades, some projects based on (JIT) compiler generation and VM frameworks appeared.

### 6.1 JIT Compiler Generation

**Pypy and Truffle** The PyPy project generates a JIT Compiler from an interpreter written in RPython and based on meta-tracing [10, 43, 3]. It is used not only to develop a Python VM but also for other languages [11, 13, 35] Truffle is a language implementation framework written in Java on top of Graal Compiler [50, 52, 18]. It is based on a self-optimizing AST interpreter and JIT compilation via partial evaluation. There are implementations of many languages implemented using this approach [40, 35, 51, 45].

Both Truffle and Pypy are engineered to generate optimizing JIT compilers. In contrast, our goal is to generate a baseline JIT compiler to reduce warm-up time. Our meta-compiler knows which expressions in the interpreter code are computable at JIT compile time, thus generating staged optimizations. The code generated by our meta-compiler is installed with the remaining VM code, allowing developers to read, modify, or mix generated and handwritten code before the VM deployment.

Another difference between these approaches is the compilation and meta-compilation granularity. Truffle performs meta-compilation at runtime on a method level. Pypy performs meta-compilation ahead of time to generate a tracing interpreter. This tracing interpreter is executed at runtime to perform JIT compilation at a trace granularity. Druid performs meta-compilation ahead of time with an instruction handler granularity. Implementation-wise, since handlers are methods, Druid is a source-to-source method-level meta-compiler, translating interpreter bytecode handlers into JIT compiler bytecode handlers.

**Ahead of time meta-compilation** Latifi *et al.,* propose to accelerate the IR generation of Truffle-based JIT compilers using the Graal Compiler ahead of time to pre-compute IR generators [32]. Their approach is based on the second Futamura projection [23].

This work is the closest to ours that we can identify. They generate a JIT compiler ahead of time without runtime information, like us. The fact that they generate IR to accelerate the optimizing JIT compiler, while we generate a baseline one, makes both approaches complementary in a VM with multiple execution tiers.

Another difference between the approaches is that we developed directly the meta-compiler without the need for partial evaluation. Pros and cons between these approaches are studied in [9].

**Targetting compiler backends** Different authors propose to generate code targetting dynamic low-level optimizing engines like VCODE [20], Dynamo [5], or SPUR [6]. This is similar to our approach based on targeting Cogit. We target Cogit not only because of its machine code generation but also for the JIT compiler infrastructure support (*i.e.,* hot-spot detection, deoptimization, PICs).

### 6.2 Compiler Generation

In 1971, Futamura proposed to use partial evaluation to generate compilers via self-application [23]. Since then, many authors have followed that approach for various programming languages [7, 30]. A recent example is LuaAOT, a compiler for Lua derived from the interpreter [39]. Other authors propose a direct approach to generate compilers by writing program generator generators (Cogen) [9, 33].

These solutions generate a compiler from a language definition (*e.g.,* an interpreter) and compilation hints (*e.g.,* annotations), like our approach using intrinsics in the interpreter to guide meta-compilation. The key distinction lies in the kind of the generated compiler: our JIT compiler must interact with an additional runtime tier (*i.e.,* hot-spot detection, deoptimizations), meanwhile, a standard compiler does not.

### 6.3 VM Frameworks

There are related works on how to build Virtual Machines efficiently. Most of them, including ours, share the idea of decoupling the language definition from VM optimizations.

**Efficient interpreters generation** Ertl, Gregg *et al.,* are responsible for the Vmgen project, a framework to generate efficient interpreters in C from a declarative definition [22, 14]. They have built a DSL to define the language semantics and apply staged optimizations to the generated interpreter (*e.g.,* super-instructions).

While the goal of both projects is indeed different, (they generate interpreters and we generate baseline JIT compilers from interpreters), their syntax to define optimizations is analogous to our guiding intrinsics. Both approaches are complementary and help VM developers build efficient execution engines in a declarative way.

**Other approaches** Yermolovich *et al.,*. have built a VM for LUA based on a hierarchical layering architecture of a host VM running a guest VM [53]. They optimize the guest VM using a tracing JIT compiler based on interpreter hints in the host VM. Geoffray, Thomas *et al.,* developed VMKit to have a common substrate that eases the development of high-level VMs [24].

These approaches are architectural and based on component reuse, ours is a generative approach. Druid meta-compilation happens ahead of time, targeting Cogit [36], the JIT compiler backend defined on Slang.

## 7 Conclusion

In this paper, we presented Druid, a meta-compiler generating baseline JITs from interpreters. Druid introduces interpreter annotations and intrinsics that guide meta-compilation. Druid targets Cogit, a JIT compiler backend used in the Pharo VM. Our solution stages computations to be executed at JIT compile time, and finally lowers code to Cogit's register transfer language, generating a JIT compiler frontend.

Our generated front end improves interpreter performance by 2× on average and achieves 0.70× the performance of the existing JIT compiler frontend that has been manually hand-tuned for 10 years, without introducing significative JIT compile time overheads. We show that meta-compiling the Pharo interpreter required only 60 semantic-preserving changes in the interpreter, mostly refactorings and code annotations, and VM maintenance is know decoupled between the implementors of a particular language and the implementors of the compiler infrastructure. In the future, we plan to explore the ahead-of-time meta-compilation of peephole optimizations, *static type prediction* and *dynamic super-instructions* for comparisons and jumps.

**Acknowledgements** This work was funded by Inria's Action Exploratoire AlaMVic.

## A Instrinsics and Annotations for Meta-compilation

This appendix presents the intrinsic functions and annotations supported by Druid for the baseline JIT compiler generation in the Pharo VM.

### A.1 Instrinsics

We perform inlinings of invoked functions during the abstract interpretation of the VM instructions defined in the interpreter (*i.e.,* meta-interpretation). However, there are functions in the interpreter in which our meta-compiler has defined functions for their interpretation, these are *instrinsic* (c.f., Section 3.2).

Tables 3 - 7 describe all the intrinsics used in Druid for the baseline JIT compiler generation in the Pharo VM. Intrinsic's implementations are defined in the meta-compiler to directly manipulate the Druid IR (control flow graph).

Figure 14 shows the implementation of `ifTrue:` intrinsic, where a conditional branch is inserted in the IR.

**Table 3** Guiding Intrinsics used during meta-compilation.

| Function | Description |
| --- | --- |
| | **Druid** |
| druidIgnore: | Ignore code during meta-interpretation. |
| interpreterIgnore: | Ignore code by the interpreter but meta-interpreted. |
| druidStageable: | Mark code as stageable for meta-compilation. |
| druidForceInterpretation | Jump the execution to the interpreter (deoptimization). |

**Table 4** Essential Intrinsics used during meta-compilation.

| Function(s) | Description |
| --- | --- |
| | **Control flow** |
| value value: cull: | Block closure activation. |
| ifTrue: caseOf:otherwise: | Conditional control flow. |
| whileTrue whileTrue: | Condition control flow for loops. |
| | **Comparisons** |
| = ~= < <= > >= \| & | Operators for equality, inequality, less, less or equal, greater, greater or equals, or and and. |
| | **Arithmetics** |
| + - * / // \ negated | Operators for addition, subtraction, multiplication, division, integer division, mod and negation. |
| | **Bit Manipulation** |
| << >> >>> bitAnd: bitOr: bitXor: anyMask: rotateLeft:by: | Operators and functions for left shift, right shift, arithmetic shift, and, or, xor, mask comparison and rotation. |

**Table 5** VM-specific Intrinsics used during meta-compilation.

| Function(s) | Description |
|---|---|
| | **SmallIntegers** |
| sumSmallInteger:with: ifOverflow: subSmallInteger:with: ifOverflow: multiplySmallInteger:with: ifOverflow: | SmallInteger object operations for addition, subtraction and multiplication with custom overflow check. |
| | **Stack access** |
| stackTop stackValue: | Read values from the stack. |
| pop: push: pop:thenPush: | Pop or push values from the stack. |
| | **Memory access** |
| byteAt:put: longAt: longAt:put: | Load and write signed integers in 8, 32 and 64 bits. |
| uint8At: uint8At:put: uint16At: uint16At:put: uint32At: uint32At:put: uint64At: uint64At:put: | Load and write unsigned integers in 8, 16, 32 and 64 bits. |

**Table 6** VM-specific Intrinsics used during meta-compilation (continuation).

| Function | Description |
|---|---|
| | **Message send** |
| normalLiteralSelectorAt: argumentCount: | Message send of a literal selector. |
| normalSendSplSelector: argumentCount: | Message send of a special selector. |
| sendSuper:numArgs: sendDirectSuper:numArgs: | Super message sends. |
| | **Method invocation** |
| iframeNumArgs: | Access to the number of arguments. |
| iframeMethod: | Access to the method object. |
| itemporary:in: itemporary:in:put: | Load or store a value from a temporary variable. |
| fetchNextBytecode | Load the next bytecode to execute. |
| fetchByte | Load the next byte in the bytecode sequence. |
| commonReturn: commonCallerReturn: | Return from a method or closure activation. |
| internalMustBeBoolean: | Throws a *MustBeBoolean* exception. |
| checkEventsCnxtSwitch: | Execution interruptions (GC and Threads management). |
| | **Pharo features** |
| error: assert: deny: | VM errors and assertions. |
| createFullClosureInIndex: numCopied:ignoreContext: | Instantiate a Block closure object. |
| executeFullCogBlock: closure:mayContextSwitch: | Activate a Block closure compiled to machine code. |
| remember: | Adds an object in the Remembered Set [49]. |
| newHashBitsOf: | Calls a function to create a hash number for an object. |

**Table 7** Instrinsic variables in the meta-compiler.

| Variable | Description |
| --- | --- |
| | **Execution** |
| instructionPointer | Stores the current instruction pointer. |
| framePointer | Stores the current frame pointer. |
| stackPointer | Stores the current stack pointer. |
| stackLimit | Stores the limit of the stack. |
| argumentCount | Stores the argument count for the current method. |
| primFailCode | Stores the current fail code of a primitive execution. |
| classTableFirstPage | Pointer to the first page of the class table. |
| | **Well known objects** |
| nilObj trueObj falseObj | Pointers to nil, true and false objects. |
| hiddenRootsObj | Pointer the root objects table. |
| | **Garbage collection** |
| scavengeThreshold | Pointer to the threshold for Scavenger. |
| freeStart | Pointer to the beginning of a free bunch of memory. |
| newSpaceStart | Pointer to the beginning of the *New Space*. |
| spaceMaskToUse newSpaceMask oldSpaceMask permSpaceMask | Stores a mask to know if an object is in the *New*, *Old* or *Permanent Space*. |
| | **Extension bytecodes** |
| extA extB | Stores values for extended bytecodes. |
| numExtA numExtB | Stores the number of extended bytecodes executed consecutively. |

### A.2 Annotations

Druid uses *annotations* to declare meta-data for the VM instructions in the interpreter. The meta-data are analyzed during meta-compilation to guide the meta-interpretation and to have extra information required by the JIT compiler framework.

Table 8 describes all the annotations supported by Druid. Figure 15 presents an extract of the IR generators bytecode table for the JIT compiler generated by Druid using extra meta-data from annotations. The JIT compiler framework uses this information during the JIT compilation of methods on runtime.

## B Examples of Meta-compilation Results

In this section, we present some examples of VM instructions in the Interpreter and the version for a baseline JIT compiler meta-compiled by Druid as it is explained in

```
interpretIfTrueWith: aRBMessageNode

  | conditionalJump trueBranchBasicBlockOut startingBasicBlock executionStateBeforeBranch |
  "Condition"
  aRBMessageNode receiver acceptVisitor: self.
  conditionalJump := self
    instantiateNoResultInstruction: DRBranchIfCondition
    operands: { DREqualsThanComparison new. self popOperand. true asDRValue }.
  self currentBasicBlock endInstruction: conditionalJump.

  "Branch"
  executionStateBeforeBranch := executionState copy.
  trueBranchBasicBlockOut := self
    branchFrom: self currentBasicBlock
    onEdge: [ :branchEntryBlock | conditionalJump trueBranch: branchEntryBlock ]
    doing: [ self interpretBlockBody: aRBMessageNode arguments first ].

  "Merge point"
  startingBasicBlock := self currentBasicBlock.
  self newBasicBlock.
  self currentBasicBlock addPredecessor: startingBasicBlock.
  conditionalJump falseBranch: self currentBasicBlock.

  "Could happen that the evaluated block had a non local return.
  In that case, this block should not arrive to this merge point
    => do not add the jump"
  trueBranchBasicBlockOut hasFinalInstruction
    ifTrue: [ executionState := executionStateBeforeBranch ]
    ifFalse: [ executionState := self
        onBlock: currentBasicBlock
        mergeAll: { executionStateBeforeBranch. executionState copy }.
      trueBranchBasicBlockOut jumpTo: self currentBasicBlock ]
```

**Figure 14** Meta-interpretation of ifTrue: intrinsic.

Section 3 and using the intrinsics and metadata listed in Appendix A. Since the VM instructions in the PharoVM are too complex, we present simplified versions of them to show how Druid meta-compiles the main language features.

### B.1 Meta-compilation of Primitives

Figure 16 presents the primitive to make the addition between two SmallInteger objects (receiver and argument). At left is the definition in the interpreter and at right is the meta-compiled version of that definition for the JIT compiler frontend.

The annotation numberOfArguments: defines which values are accessed by stackValue: intrinsic. It is required because, by the calling convention in the machine code, arguments (and the receiver) are passed by registers instead of the stack. Thus, values accessed by the stack are replaced by expected registers (*i.e.,* ReceiverResultReg, Arg0Reg) during meta-compilation.

**Table 8** Annotations with meta-compilation options.

| Annotation | Description |
|---|---|
| **For primitives** | |
| numberOfArguments: | One or many expected numbers of arguments for the primitive method, failing JIT compilation in other cases. |
| customisedReceiverFor: | Conditional JIT compilation based on primitive method class. |
| druidExitPoint | Primitive fails at this point on runtime. |
| **For bytecodes** | |
| compilationInfo: | Adds extra information on bytecode declaration for the JIT compiler framework (*i.e.,* mappings between bytecodes and JITted code, jumps, etc). |
| druidInfo: | Computes extra information during bytecode meta-compilation (*i.e.,* message send invocations). |
| needsFrameNever:<br>needsFrameIfInBlock:<br>needsFrameIfImmutability: | Says if the bytecode has access to the frame (if none, the frame is not created in JITted code). |

The annotation customisedReceiverFor: indicates that this primitive only works if the receiver is an instance of a special class, SmallInteger in this case. The meta-compiler optimizes it by assuming that the primitive method must be installed in the expected class and makes that check at JIT compile time (*i.e.,* mclassIsSmallInteger) to avoid compilation in other cases. Several compiler optimizations remove the check of the receiver type at runtime, only the check for the argument is performed in JIT-compiled code (by inlining the isIntegerObject: function).

Finally, the intrinsic sumSmallInteger:with:ifOverflow: is meta-compiled to deal with tag, perform the addition and check overflows. Meta-compilation of the intrinsic pop:thenPush: moves the result to the expected register (*i.e.,* ReceiverResultReg), following the register-based calling convention. In cases where the primitive fails (*i.e.,* primitiveFail), the flow jumps to the end of the generated machine code, where a fallback method will be executed.

### B.2 Meta-compilation of Bytecodes

Figure 17 shows the input (left) and output (right) of the meta-compilation of a bytecode responsive to push a receiver instance variable value to the stack.

The interpreter's definition uses the currentBytecode variable to compute the instance variable index. This variable contains the executed bytecode index in the bytecode table, thus the same definition is used in many entries for accessing different indexes. We meta-compile different definitions for each entry by staging the currentBytecode to the current number and constant-fold it, if possible. The example

```
JITFrontend >> bytecodeTable

 ^ {
   ...
   { 1. 64. 64. #gen_PushTemporaryVariableBytecode0 }.
   ...
   { 1. 88. 88. #gen_ReturnReceiver. #return. #isMappedInBlock. #needsFrameIfInBlock:. 0 }.
   ...
   { 1. 92. 92. #gen_ReturnTopFromMethod. #return. #isMappedInBlock. #needsFrameIfInBlock:.
   ↪   -1 }.
   ...
   { 1. 95. 95. #gen_ExtNopBytecode. #needsFrameNever:. 0 }.
   { 1. 96. 96. #gen_BytecodePrimAdd. #isMapped }.
   ...
   { 1. 128. 128. #gen_SendLiteralSelector0ArgsBytecode0. #isMapped }.
   ...
   { 1. 176. 176. #gen_ShortUnconditionalJump0. #branch. #v3:ShortForward:Branch:Distance: }.
   ...
   { 1. 184. 184. #gen_ShortConditionalJumpTrue0. #branch. #isBranchTrue. #isMapped. #v3:
   ↪  ShortForward:Branch:Distance: }.
   ...
   { 1. 200. 200. #gen_StoreAndPopReceiverVariableBytecode0. #isInstVarRef.
   ↪  #isMappedIfImmutability. #needsFrameIfImmutability:. -1 }.
   ...
   { 3. 255. 255. #unknownBytecode } }
```

**Figure 15** Extract of the bytecode table generated by Druid using meta-data.

in Figure 17 was meta-compiled for index = 1 in the bytecode table, so it accesses the second instance variable (0-based). This value was constant folding, after the inline of fetchPointer:ofObject: function, to offset = 16 (bytes) in the meta-compiled code (8 bytes of the object header + 8 bytes of the first instance variable).

The way of accessing the receiver object from the frame in the interpreter is meta-compiled by using the register-based calling convention (*i.e.,* using ReceiverResultReg). Since previous bytecodes in the compiling method could override the register value, the definition requires calling ensureReceiverResultRegContainsSelf compiler function.

Bytecode meta-compilation implies dynamic register allocation at JIT-compile time by using allocateRegNotConflictingWith:ifNone: compiler function. Each meta-compiled bytecode definition keeps a variable (live) with the status of already used registers to avoid conflicts.

To push the value, Druid uses the ssPushBase:offset: function exposed by the compiler backend to track the values in the stack and perform peep-hole optimizations during JIT compilation.

In the end, the fetchNextBytecode intrinsic, present in every bytecode definition in the interpreter, is ignored during meta-compilation due to the JIT compiler framework resolves the compilation of a sequence of bytecodes inside the method.

```
Interpreter >> primitiveAdd
  <numberOfArguments: 1>
  <customisedReceiverFor: #smallInteger>

  | rcvr arg rawInt1 rawInt2 rawResult result |
  "Get the values from the stack"
  rcvr := self stackValue: 0.
  arg := self stackValue: 1.
  "Check small integer objects"
  (memory isIntegerObject: rcvr)
    ifFalse: [ ^ self primitiveFail ].
  (memory isIntegerObject: arg)
    ifFalse: [ ^ self primitiveFail ].
  "Perform the addition, checking overflow"
  result := self
    sumSmallInteger: rcvr
    with: arg
    ifOverflow: [ ^ self primitiveFail ].
  "Pop values and push the result"
  self pop: 2 thenPush: result
```

```
JITFrontend >> gen_PrimitiveAdd
  "AutoGenerated by Druid"

  | jump1 jump2 currentBlock |
  self mclassIsSmallInteger ifFalse: [ ^
     ↪ UnimplementedPrimitive ].
  self TstCq: 1 R: Arg0Reg.
  jump1 := self JumpZero: 0.
  self MoveR: Arg0Reg R: TempReg.
  self AddCq:  -1 R: TempReg.
  self MoveR: ReceiverResultReg R: ClassReg.
  self AddR: ClassReg R: TempReg.
  jump2 := self JumpOverflow: 0.
  self MoveR: TempReg R: ReceiverResultReg.
  self genPrimReturn.
  currentBlock := self Label.
  jump1 jmpTarget: currentBlock.
  jump2 jmpTarget: currentBlock.
  ^ CompletePrimitive
```

**Figure 16** Meta-compilation result of addition primitive for SmallInteger objects.

```
Interpreter >> bytecodePushReceiverVariable

  | fieldIndex receiver |
  fieldIndex := currentBytecode bitAnd: 16rF.
  receiver := stackPages longAt: framePointer +
     ↪ FoxIFReceiver.
  self push: (objectMemory fetchPointer:
     ↪ fieldIndex ofObject: receiver).
  self fetchNextBytecode
```

```
JITFrontend >>
     ↪ gen_bytecodePushReceiverVariable_1
  "AutoGenerated by Druid"

  | live currentBlock t0 |
  live := 0.
  self ensureReceiverResultRegContainsSelf.
  t0 := self
        allocateRegNotConflictingWith: live
        ifNone: [ ^ self unknownBytecode ].
  live := live bitOr: (self registerMaskFor: t0).
  self MoveR: ReceiverResultReg R: t0.
  self ssPushBase: t0 offset: 16.
  ^ 0
```

**Figure 17** Push receiver variable bytecode meta-compilation with index = 1.

### B.3 Using the JIT Compiler Framework

Figure 18 introduces a conditional jump bytecode for short distances. The offset on the interpreter is computed based on the currentBytecode (the bytecode index in the table). Similar to the previously presented case, we meta-compile a version for each entry in the table (with offset = 1 in the given example).

```
Interpreter >> shortConditionalJumpTrue
  <compilationInfo: #isMapped>
  <compilationInfo: #branch>
  <compilationInfo: #isBranchTrue>

  | boolean offset |
  offset := (currentBytecode bitAnd: 7) + 1.
  boolean := self stackTop.
  self pop: 1.
  boolean = objectMemory trueObject
    ifTrue: [ self jump: offset ]
    ifFalse: [
      boolean = objectMemory falseObject
    ↪ ifFalse: [
        self internalMustBeBoolean: boolean ].
      self fetchNextBytecode ]
```

```
JITFrontend >>
    ↪ gen_ShortConditionalJumpTrue0
  "AutoGenerated by Druid"

  | jump1 s5 s2 currentBlock s12 t0 jump2 s9 |
  self annotateBytecode: self Label.
  t0 := ... "Allocate register"
  (self ssValue: 0) copyToReg: t0.
  self ssPop: 1.
  s5 := objectMemory trueObject.
  self ssFlushStack.
  self CmpCq: s5 R: t0.
  jump1 := self JumpNonZero: 0.
  s9 := bytecodePC + 2.
  self Jump: (self ensureFixupAt: s9).
  jump2 := self Jump: 0.
  currentBlock := self Label.
  jump1 jmpTarget: currentBlock.
  s12 := objectMemory falseObject.
  self CmpCq: s12 R: t0.
  jump1 := self JumpZero: 0.
  self MoveR: t0 R: TempReg.
  self CallRT:
    ↪ ceSendMustBeBooleanTrampoline.
  currentBlock := self Label.
  jump2 jmpTarget: currentBlock.
  jump1 jmpTarget: currentBlock.
  ^ 0
```

**Figure 18** Conditional true jump bytecode meta-compilation with offset = 1.

The annotation with the compilation info isMapped in the interpreter definition indicates that this bytecode needs to save the link between the bytecode index and the address of the JIT-compiled code because of deoptimization possibilities. This link is created using the annotateBytecode: function, exposed by the JIT compiler framework, at the beginning of the JITFrontend definition. The deoptimization case occurs when the internalMustBeBoolean: intrinsic is executed.[3] As this case produces an exception and escapes from the fast path that we want to JIT compile, this intrinsic is meta-compiled as a call to the trampoline ceSendMustBeBooleanTrampoline to manage the case by the runtime system. The trampolines are generated by the JIT compiler framework at the beginning of execution and deal with the required transformation to continue the execution in the runtime (*e.g.,* deoptimize the stack frame).

[3] In Pharo, conditionals are written as messages to boolean objects (*e.g.,* boolean ifTrue: [ ... ]). The message is compiled to bytecode as a conditional jump. The bytecode must support the case of a non-boolean object as the receiver (error case).

The meta-compiled generator handler definition in the JIT frontend uses functions exposed by the JIT backend to perform optimizations at JIT compile time. On one side, the functions to access and modify the stack, ssValue: and ssPop:, use an abstract stack to track operations and perform peep-hole optimizations. As a result, we must call ssFlushStack before any potential deoptimization to ensure that values in registers are spilled to the stack in case of continuing the execution in the runtime. Conversely, we use the ensureFixupAt: function to track jumps between bytecodes and perform dead code elimination.

The JITFrontend version in Figure 18 also presents staged operations at JIT compile time. In the expression where the bytecode to jump is computed, s9 := bytecodePC + 2, our meta-compiler identifies which values can be *constant folded* at meta-compile time—the final value 2 is produced by 1 (offset) + 1 (next bytecode is always avoided)—and which values must be staged for JIT compile time—the addition bytecodePC + 2 is evaluated by the JIT compiler since it does not rely on runtime information.

A similar case occurs on comparing boolean objects—the expressions objectMemory trueObject and objectMemory falseObject return a pointer to each object. Since they are special objects, they are fixed in memory and are not moved during Garbage Collection. Thus, we cannot get the value at meta-compile time but at JIT compile time and inline the value in the JIT-compiled code.

### B.4 Message Send

Figure 19 shows both definitions for a *message send* bytecode. The presented bytecode is a special case for = message, used for equality comparison in Pharo.

```
Interpreter >> bytecodeSendEquals
  <compilationInfo: #isMapped>
  <druidInfo: #hasSend>

  self druidIgnore: [
    | rcvr arg |
    rcvr := self stackValue: 1.
    arg := self stackValue: 0.
    (objectMemory areIntegers: rcvr and: arg)
      ↪ ifTrue: [
      ^ self booleanCheat: rcvr = arg ] ].

  self normalSendSpecialSelector: 6
      ↪ argumentCount: 1
```

```
JITFrontend >> gen_bytecodeSendEquals
  "AutoGenerated by Druid"

  self marshallSendArguments: 1.
  self
    genMarshalledSend: -7
    numArgs: 1
    sendTable: ordinarySendTrampolines.
  ^ 0
```

**Figure 19** Special message send bytecode meta-compilation for = message.

The interpreter definition of the bytecode includes the two optimizations commented in Section 5.4. A *static type prediction* is implemented for the case of SmallInteger objects, performing the comparison of the tagged values. Also, *dynamic superinstructions* is implemented inside the booleanCheat: function, where a look-ahead of the next bytecode is performed to check if it is a conditional jump and directly execute it with the computed boolean value. Our meta-compiler does not support these optimizations yet, so they are ignored using the intrinsic druidIgnore:. As we say in Section 7, we are still working to support them.

The targeting JIT compiler backend Cogit already has support for message sends. We meta-compile the bytecodes with message sends by using marshallSendArguments: and genMarshalledSend:numArgs:sendTable: JIT compiler backend functions. These functions generate calls to trampolines to perform the lookup and patch the code with PICs. They also create the mapping between bytecode and compiled code, so the annotateBytecode: function is not invoked by the frontend.

### B.5 Staging Conditional Branches

Figure 20 presents a version of the push new array bytecode, *i.e.,* it instantiates a new array object and fills all slots with references to nil object. The Interpreter version was simplified to focus on the staging features of the meta-compiler. The JITFrontend version only shows the important parts of this section.

```
Interpreter >> bytecodePushNewArray

  | size array |
  size := self fetchByte.
  size < 0 ifTrue: [
    self druidForceInterpretation.
    "..."
  ].
  size := size bitAnd: 127.
  self fetchNextBytecode.
  array := ... "Instantiate new array"
  0 to: size - 1 do: [ :i |
    objectMemory
      storePointerUnchecked: i
      ofObject: array
      withValue: objectMemory nilObject ].
  self push: array
```

```
JITFrontend >> gen_bytecodePushNewArray
  "AutoGenerated by Druid"

  | ... |
  s2 := byte1.
  self ssFlushStack.
  s2 < 0 ifTrue: [
    self deoptimize.
    ^ 0 ].
  s7 := s2 bitAnd: 127.
  t0 := ... "Allocate register"
  t1 := ... "Allocate register"
  ... "Store new array in memory, saved in t1"
  s28 := s7 - 1.
  s29 := 0.
  t2 := ... "Allocate register"
  [ ((s29<=s28)) ] whileTrue: [
    self genMoveConstant: objectMemory
      ↪ nilObject R: t0.
    s34 := s29 << 3.
    self MoveR: t1 R: t2.
    self AddCq: s34 R: t2.
    self MoveR: t0 M64: 8 r: t2.
    s38 := s29 + 1.
    s29 := s38 ].
  self ssPushRegister: t1.
  ^ 0
```

■ **Figure 20** Push new array bytecode meta-compilation (some parts of the code).

In previous examples, we presented that our meta-compiler stages expressions to be executed at JIT compile time. For the cases where these expressions are the condition of a conditional branch, the meta-compiler stages the branches too. We can see the staged branches by using `ifTrue:` and `whileTrue:` functions in the JITFrontend version in Figure 20.

The first case is where the condition `size < 0 ifTrue:` in the Interpreter is meta-compiled as `s2 < 0 ifTrue:` in the JITFrontend. Since the value of `size` is encoded in the 2 bytes instruction, the condition does not depend on runtime information and can be resolved at JIT compile time. In this case, the JIT compiler backend already performs the fetch of the next bytes storing them in variables for the frontend, so the `fetchByte` intrinsic is meta-compiled as an access to the expected variable `byte1`. Inside the branch, the `druidForceInterpretation` intrinsic is meta-compiled by calling `deoptimize` JIT compiler backend function. In that case, a trampoline is called to continue the execution on the Interpreter (slow path).

The second case of staged branches occurs during the meta-compilation of the `to:do:` function, defined based on `whileTrue:` intrinsic. Here also, the condition depends only on the `size` variable, thus the loop is staged. In this case, the meta-compiler must determine which expressions belong to the pre-header, the conditional header, and the loop body.

## About the authors

**Nahuel Palumbo** nahuel.palumbo@inria.fr
I am a Phd student at INRIA - EVREF team working on meta-compilation and virtual machines for object-oriented programming.
https://orcid.org/0009-0001-5004-5632

**Guillermo Polito** guillermo.polito@inria.fr
I am an Inria Researcher at the CRIStAL laboratory, where I am expert in software engineering and language implementations, with a particular focus on test generation. I work in the Evref team from Inria Lille-Nord Europe.
https://orcid.org/0000-0003-0813-8584

**Stéphane Ducasse** stephane.ducasse@inria.fr
I'm an expert in object design, object language design, reflective programming, and the maintenance and evolution of large applications (visualization, metrics, meta-modeling). My work on traits has been introduced in AmbientTalk, Slate, Pharo, Perl-6, PHP 5.4 and Squeak. They have been ported to JavaScript. It has influenced the Scala and Fortress languages. I'm one of the founders of Pharo, a new pure open-source object language inspired by Smalltalk. I head its industry consortium. I'm one of the designers of Moose, an analysis platform. I was one of the founders of Synectique, a company offering dedicated analysis tools.
https://orcid.org/0000-0001-6070-6599

**Pablo Tesone** pablo.tesone@inria.fr
Pablo Tesone is Ingénieur de recherche of the Pharo Consortium at INRIA University of Lille, within the Evref team, with more than 10 years of experience in industrial projects. He has done a PhD on Dynamic Software Update applied to Live programming environments, distributed systems and robotic applications. He is interested in improving development tools and the daily development process. He is an enthusiast of the object oriented programming and their tools. He collaborates with different open source projects like the ones in the Pharo Community and the Uqbar Foundation.
https://orcid.org/0000-0002-5615-6691